\documentclass[nofootinbib,preprintnumbers]{revtex4}

\usepackage{epsfig}
\newcommand{\beq}{\begin{equation}}
\newcommand{\eeq}{\end{equation}}
\def\bea{\begin{eqnarray}}
\def\eea{\end{eqnarray}}

\begin{document}
\title{Scaling of the Higgs Self-coupling and bounds on the Extra Dimension}

\author{A.~S.~Cornell}
\email[Email: ]{alan.cornell@wits.ac.za}
\affiliation{National Institute for Theoretical Physics; School of Physics, University of the Witwatersrand, Wits 2050, South Africa}
\author{Lu-Xin~Liu}
\email[Email: ]{luxin.liu9@gmail.com}
\affiliation{National Institute for Theoretical Physics; School of Physics, University of the Witwatersrand, Wits 2050, South Africa}

\begin{abstract}
In this paper we study the one-loop evolution equation of the Higgs quartic coupling $\lambda$ in the minimal Universal Extra Dimension model, and find that there are certain bounds on the extra dimension due to the singularity and vacuum stability conditions of the Higgs sector. In the range $250GeV \sim {R^{ - 1}} \sim 80TeV$ of the compactification radius, we notice that for a given initial value $\lambda ({M_Z})$, there is an upper limit on ${R^{ - 1}}$ for a Higgs mass of $183GeV \sim {m_H}({M_Z}) \sim 187GeV$; where any other compactification scales beyond that have been ruled out for theories where the evolution of $\lambda$ does not develop a Landau pole and become divergent in the whole range (that is, from the electroweak scale to the unification scale). Likewise, in the range of the Higgs mass $152GeV \sim {m_H}({M_Z}) \sim 157GeV$, for an initial value $\lambda ({M_Z})$, we are led to a lower limit on ${R^{ - 1}}$; any other compactification scales below that will be ruled out for theories where the evolution of $\lambda$ does not become negative and destabilize the vacuum between the electroweak scale and the unification scale. For a Higgs mass in the range $157GeV < {m_H}({M_Z}) < 183GeV$, the evolution of $\lambda$ is finite and the theory is valid in the whole range (from the electroweak scale to the unification scale) for $250GeV \sim {R^{ - 1}} \sim 80TeV$.
\end{abstract}

\date{5 May, 2011}
\preprint{WITS-CTP-71}
\maketitle


\section{Introduction}\label{sec:1}

\par The Standard Model (SM) is the gauge theory with gauge symmetry group $S{U_C}(3) \times S{U_L}(2) \times {U_Y}(1)$, and that provides a very precise description of microscopic interactions. Understanding the mechanism that breaks electroweak symmetry and generates the masses of all known elementary particles is one of the most fundamental problems in particle physics. The Higgs mechanism gives us a self-interacting scalar field which is arranged such that the neutral component of the scalar doublet acquires a vacuum expectation value which sets the scale of electroweak symmetry breaking. As a result, it provides the weak gauge bosons with masses through the absorption of the charged and neutral Goldstone bosons as their longitudinal components.

\par The speculations of the Higgs particle's interactions, and its discovery, are one of the most exciting topics in contemporary particle physics \cite{Wells}.  One of the unanswered questions about the Higgs particle is to understand the behavior of the quartic coupling $\lambda$, through which the mass of the Higgs particle, ${m_H}$, is obtained. In fact, in the SM, the Higgs boson mass is given by ${m_H} = \sqrt {\lambda} v$, where $\lambda$ is the Higgs self-coupling parameter and $v$ is the vacuum expectation value of the Higgs field: $v = {(\sqrt 2 {G_F})^{ - 1/2}} = 246GeV$ is fixed by the Fermi coupling $G_F$. Since $\lambda$ is presently unknown, the value of the SM Higgs boson mass $m_H$ cannot be derived directly.

\par The Higgs sector of the SM has two important parameters, the Higgs mass and the new physics scale $\Lambda$. Below that scale the SM is an extremely successful effective field theory that has emerged from the electroweak precision tests of the last decades. Above that scale the SM is no longer valid and must be embedded into some more general theory, the possibilities of this theory spawning a wealth of new physics phenomena. The value of $m_H$ itself can provide an important constraint to the scale to which the SM remains successful as an effective theory. Arguments can place approximate upper and lower bounds on $m_H$ itself, for example there is an upper bound based on the perturbativity of the theory up to the scale at which the SM breaks down, and a lower bound derived from the stability of the Higgs potential. If $m_H$ is too large, then the Higgs self-coupling diverges at some scale $\Lambda$, this is called the Landau pole. On the other hand, from the requirement that the scalar potential energy of the vacuum be bounded from below, the quartic coupling $\lambda$ should be positive at any energy scale. If $m_H$ is too small, $\lambda$ becomes negative at certain energy scales, at which point the Higgs potential is destabilized. The presence of the singularity and zero values for the evolution of $\lambda$ simply leads to an upper bound and a lower bound for its initial value, and new energy scales can thus be introduced and which lead to the emergence of new physics.

\par With the Large Hadron Collider (LHC) now up and running, physicists have begun to explore the realm of new physics that may operate at the TeV scale. Among these, models with extra spatial dimensions may be revealed in higher energy collider experiments. In particular, the Universal Extra Dimension (UED) model makes an interesting $TeV$ scale physics scenario which features a tower of Kaluza-Klein (KK) states for each of the SM fields, all of which have full access to the extended spacetime manifold\cite{Liu:2011gr,Cornell:2010sz,Bhattacharyya:2006ym}.  It is well known that models with extra dimensions may bring down the unification scale to a much lower energy scale \cite{Dienes:1998vg}. Therefore, instead of assuming the Renormalisation Group Equations (RGE) go from the $M_Z$ scale up to the Grand Unification Theory (GUT) scale ($10^{14} GeV$) by using the $S{U_C}(3) \times S{U_L}(2) \times {U_Y}(1)$ symmetry, the evolution of physics under the context of a UED model would be significantly different due to the modified beta functions. In the current context we will focus on the evolution of the Higgs self-coupling and explore its behaviours and correlation with the compactified extra dimension.

\par As such, in this paper, we consider a UED model with a single compactified extra dimension with an ${S_1}/{Z_2}$ symmetry. Recall that in order to explore physics at a high energy scale we use RGE as a probe to study the momentum dependence of physical quantities. Thus, in Section \ref{sec:2}, we first start from the one-loop diagrams of the contributions from the relevant KK modes to the beta function of the Higgs' quartic coupling. The evolution of the quartic coupling is then derived using the anomalous dimensions of the wave function and proper vertex renormalisation of the scalar field. In Section \ref{sec:3} the evolution equation of $\lambda$ is found to be of the Riccati type, which we then show how to solve explicitly. The position of the Landau pole and zero of the quartic coupling $\lambda$ are given in terms of different compactification radii. We find that the function $\lambda (t)$ has a Landau singularity at a relatively low energy scale as compared with the case of the SM. A very precise analysis of the position where $\lambda$ vanishes is also performed, and the new physics scale can be identified as the value of the scale at which $\lambda$ crosses zero. In Section \ref{sec:4} we quantitatively analyze the evolution of the scalar coupling from the electroweak scale up to the unification scale and exploit its evolutionary behaviour for different compactification radii $R$, where it is most interesting to investigate how $\lambda$ depends on its initial values, as well as the compactification radius. Assuming the theory is valid and consistent in the whole range (from the electroweak scale up to the unification scale), $\lambda$ must be positive and cannot be singular. As a result we obtain numerical and graphical results for the behavior of $\lambda$ which leads to bounds on the compactification radius. The last section is devoted to a summary and our conclusions.


\section{The Evolution Equations}\label{sec:2}

\par The RGE are an important tool for the search of the properties of physics at different energy scales. In the SM the Higgs quartic coupling, $\lambda$, which gives the mass of the Higgs scalar is given as:
\beq
\frac{\lambda }{2}{({\Phi ^\dag }\Phi )^2}\; . \label{eqn:1}
\eeq
It is known that the renormalized coupling constant depends on the choice of the scale parameter $\mu$, where the bare constant is independent of the renormalization scale. As a result, the evolution of the Higgs quartic coupling is given by the beta function:
\beq
\mu \frac{\partial }{{\partial \mu }}\ln {\lambda ^R} = \mu \frac{\partial }{{\partial \mu }}\ln Z_\Phi ^2 - \mu \frac{\partial }{{\partial \mu }}\ln {Z_{coupling}}, \label{eqn:2}
\eeq
where $\lambda^R$ is the renormalized quartic coupling constant (we shall drop the index $R$ for the remainder of the paper), with ${Z_\Phi }$ the wave function renormalization constants related to the scalar boson, and ${Z_{coupling}}$ as the proper vertex renormalization constant. The evolution equation of $\lambda$ for the SM has been studied in various references, see \cite{Machacek:1984zw,Cheng:1973nv,Kielanowski:2003jg}. Here we shall explicitly illustrate the contributions of the UED's KK modes to this beta function and plot their effects to the evolution of the scalar coupling. For simplicity we choose to work with the minimal UED model, i.e. the extra dimension is compactified on a circle of radius $R$ with a ${Z_2}$ orbifolding, which identifies the fifth coordinate $y \to  - y$. From a 4-dimensional view point, every field will then have an infinite tower of KK modes, with the zero modes being identified as the SM state. The 5-dimensional KK expansion of the scalar field then becomes:
\beq
\Phi (x,y) = \frac{1}{{\sqrt {\pi R} }}\{ \Phi (x) + \sqrt 2 \sum\limits_{n = 1}^ \propto  {{\Phi _n}(x)\cos (\frac{{ny}}{R})} \} \; . \label{eqn:3}
\eeq
In the bulk we have the scalar boson and gauge fields interactions as:
\beq
{\ell _{Higgs}} = \int\limits_0^{\pi R} {dy} {({D_M}\Phi (x,y))^\dag }{D^M}\Phi (x,y)\; , \label{eqn:4}
\eeq
in which the kinetic term ${D_M}\Phi (x,y) = ({\partial _M} + ig_2^5{T^a}W_M^a + \frac{i}{2}g_1^5{B_M})\Phi (x,y)$, and the $5D$ gauge fields have the form ${A_M} = ({A_\mu },{A_5})$; the fifth component of the gauge bosons, ${A_5}(x,y)$, being a real scalar which does not have any zero mode, transforming in the adjoint representation of the gauge group. Also note that the $5D$ coupling constants are related to the 4-dimensional SM coupling constants up to a normalization factor, i.e. $ \displaystyle{g_i} = \frac{{g_i^5}}{{\sqrt {\pi R} }}$, and $\displaystyle\lambda  = \frac{{{\lambda ^5}}}{{\sqrt {\pi R} }}$.

\par
\begin{figure}[tb]
\begin{center}
\epsfig{file=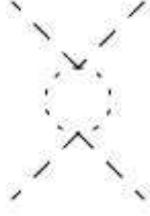,width=.15\textwidth}
\caption{\sl The one-loop corrections from the fifth component of the vector fields to the scalar coupling in Eq.(\ref{eqn:2}), introduced at each KK excited level. The dashed line is for the Higgs field, and the dotted line is for the $A_5^n$ scalar. In Table \ref{tab:1} the contributions to the anomalous dimension of the proper vertex are presented.}
\label{fig:1}
\end{center}
\end{figure}

 After integrating out the compactified dimension, the 4-dimensional effective Lagrangian has interactions involving the zero (SM) modes and the KK modes. When calculating the one-loop diagrams of the scalar quartic coupling we choose to work in the Landau gauge in what follows, as many one-loop diagrams are finite in the Landau gauge and have no contribution to the renormalization of the scalar quartic coupling. At each KK excited level the one-loop Feynman diagrams that contribute to the scalar coupling renormalization exactly mirror those of the SM field ground states \cite{Cheng:1973nv} plus new contributions from the $A_5^n$ interactions due to the fifth component of the vector fields, as shown in Fig.\ref{fig:1}, where the anomalous dimensions ${\gamma _{coupling}}$ (${\gamma _{coupling}} = \mu \frac{\partial }{{\partial \mu }}\ln {Z_{couping}}$) can be extracted from the divergent parts by using dimensional regularization. In Table \ref{tab:1} we list the results of the proper vertex anomalous dimensions of Fig.\ref{fig:1} for the $A_5^n$ field as well as those of the gauge fields $A_\mu$ in the SM.

\begin{table}[t]
\caption{\small\sl The proper vertex anomalous dimensions for $A_5^n$ in Fig.\ref{fig:1}, as well as the gauge fields $A_\mu$ in the SM, each column referring to the type of contributions related to $g_1^4, g_1^2 g_2^2$, and $g_2^4$ respectively.}
\label{tab:1}
\begin{tabular}{|c|c|c|c|c|c|}
\hline
\multicolumn{3}{c|}{$\gamma _{coupling}(UED)$} & \multicolumn{3}{c|}{$\gamma _{coupling}(SM)$} \\ \hline
$\;\; \;\; g_1^4 \;\; \;\; $ & $\;\; \;\; g_1^2 g_2^2 \;\; \;\; $ & $\;\; \;\; g_2^4 \;\; \;\; $ & $\;\; \;\; g_1^4 \;\; \;\; $ & $\;\; \;\; g_1^2 g_2^2 \;\; \;\; $ & $\;\; \;\; g_2^4 \;\; \;\; $ \\ \hline &&&&&\\
$\;\; \;\; \displaystyle -\frac{1}{4}\cdot \frac{9}{25} g_1^4\;\; \;\; $ & $\;\; \;\; \displaystyle -\frac{1}{4}\cdot \frac{6}{5} g_1^2 g_2^2\;\; \;\; $ & $\;\; \;\; \displaystyle -\frac{3}{4} g_2^4  \;\; \;\; $ & $\;\; \;\; \displaystyle -\frac{9}{4}\cdot \frac{3}{25} g_1^4\;\; \;\; $ & $\;\; \;\; \displaystyle -\frac{9}{4}\cdot\frac{2}{5} g_1^2 g_2^2 \;\; \;\; $ & $\;\; \;\; \displaystyle -\frac{9}{4} g_2^4 \;\; \;\; $ \\ &&&&& \\ \hline
\end{tabular}
\end{table}

\par For simplicity, we have omitted a common multiplicative factor of $\displaystyle\frac{1}{{16{\pi ^2}\lambda}}$ in Table \ref{tab:1}. The coupling constant ${g_1}$ is also chosen to follow the conventional $SU(5)$ normalization. Between the scale ${R^{ - 1}}$, where the first KK states are excited, and the cutoff scale, there are finite quantum corrections of the KK states to the scalar coupling. Following the discussions in \cite{Cornell:2010sz}, the one-loop evolution equation for the scalar quartic coupling from these cumulative effects of the KK modes has the following form:
\beq
16{\pi ^2}\frac{{d\lambda }}{{dt}} = \beta _\lambda ^{SM} + \beta _\lambda ^{UED}\; , \label{eqn:5}
\eeq
where the beta functions are given by
\bea
\beta _\lambda ^{SM} &=& 12{\lambda ^2} - \left( {\frac{9}{5}g_1^2 + 9g_2^2} \right)\lambda  + \frac{9}{4}\left( {\frac{3}{{25}}g_1^4 + \frac{2}{5}g_1^2g_2^2 + g_2^4} \right) + 4\lambda Tr[3Y_U^\dag {Y_U} + 3Y_D^\dag {Y_D} + Y_E^\dag {Y_E}]\nonumber \\
&& - 4Tr[3{(Y_U^\dag {Y_U})^2} + 3{(Y_D^\dag {Y_D})^2} + {(Y_E^\dag {Y_E})^2}]\; , \label{eqn:6}
\eea
and
\bea
\beta _\lambda ^{UED} &=& (S(t) - 1)\left\{ {12{\lambda ^2} - 3\left( {\frac{3}{5}g_1^2 + 3g_2^2} \right)\lambda  + \left( {\frac{9}{{25}}g_1^4 + \frac{6}{5}g_1^2g_2^2 + 3g_2^4} \right)} \right\}\nonumber\\
&& + 2(S(t) - 1)\left\{ {4\lambda Tr[3Y_U^\dag {Y_U} + 3Y_D^\dag {Y_D} + Y_E^\dag {Y_E}] - 4Tr[3{{(Y_U^\dag {Y_U})}^2} + 3{{(Y_D^\dag {Y_D})}^2} + {{(Y_E^\dag {Y_E})}^2}]} \right\}\; . \label{eqn:7}
\eea
Note that here $t = \ln (\mu /{M_Z})$ is the energy scale parameter, where we have chosen the $Z$ boson mass as the renormalization point, and $S(t) = {e^t}{M_Z}R$. In deriving $\beta _\lambda ^{UED}$, for the factor of ${g_1}^4$ as an explicit example, we have $\displaystyle-\frac{1}{4} \cdot \frac{9}{{25}}{g_1}^4$ from the $A_5^n$ contributions, together with a factor $\displaystyle-\frac{9}{4} \cdot \frac{3}{{25}}{g_1}^4$ which is read off from ${\gamma _{coupling}}$ (SM) in Table \ref{tab:1}, as a result the related one-loop graphs of the $A_\mu ^n$ mode mirrors those of the SM (zero) mode. As given in Eq. (\ref{eqn:2}), this then leads to a total factor of $\displaystyle\frac{9}{{25}}g_1^4$. The second line in $\beta _\lambda ^{UED}$ is attributed to the fermion-Higgs Yukawa couplings, where at each KK level, the KK fermions have graphs exactly mirroring the zero mode 4-dimensional SM ground state. Furthermore, the KK fermions at a given level are vector-like, this then accounts for a relative factor of 2 between the first and second line in $\beta _\lambda ^{UED}$ for the proportionality factor $S(t) - 1$ (in the Landau gauge, the other one-loop diagrams \cite{Machacek:1984zw} related to the gauge field contributions to the vertex renormalization vanish as the ${A_5}^n$ contributions associated with these diagrams are not divergent, that is, they have no contribution to the vertex renormalization constant).

\par In addition, the other physical parameters, such as Yukawa couplings and gauge couplings do not evolve in the old SM fashion. Their beta functions are as follows \cite{Liu:2011gr, Cornell:2010sz}:
\bea
16{\pi ^2}\frac{{d{f_i}^2}}{{dt}} &=& {f_i}^2[2(2S - 1)T - 2{G_U} + 3S{f_i}^2 - 3S\sum\limits_j  {{h_j }^2} {\left| {{V_{ij }}} \right|^2}],\nonumber\\
16{\pi ^2}\frac{{d{h_j }^2}}{{dt}} &=& {h_j }^2[2(2S - 1)T - 2{G_D} + 3S{h_j}^2 - 3S\sum\limits_i {{f_i}^2} {\left| {{V_{ij}}} \right|^2}],\nonumber\\
16{\pi ^2}\frac{{d{y_a}^2}}{{dt}} &=& {y_a}^2[2(2S - 1)T - 2{G_E} + 3S{y_a}^2],\; \label{eqn:8}
\eea
where $f_i^2$ and $h_j^2$ are eigenvalues of $Y_U^\dag {Y_U}$ and $Y_D^\dag {Y_D}$ respectively, and ${Y_E} = diag({y_e},{y_\mu },{y_\tau })$, $T = Tr[3Y_U^\dag {Y_U} + 3Y_D^\dag {Y_D} + Y_E^\dag {Y_E}]$, ${G_U} = 8g_3^2 + \frac{9}{4}g_2^2 + \frac{{17}}{{20}}g_1^2 + (S - 1)(\frac{{28}}{3}g_3^2 + \frac{{15}}{8}g_2^2 + \frac{{101}}{{120}}g_1^2)$, ${G_D} = 8g_3^2 + \frac{9}{4}g_2^2 + \frac{1}{4}g_1^2 + (S - 1)(\frac{{28}}{3}g_3^2 + \frac{{15}}{8}g_2^2 + \frac{{17}}{{120}}g_1^2)$ and ${G_E} = (\frac{9}{4}g_2^2 + \frac{9}{4}g_1^2) + (S - 1) (\frac{{15}}{8}g_2^2 + \frac{{99}}{{40}}g_1^2)$. The evolution of the quark flavor mixing matrix in the charged current is governed by
\bea
16{\pi ^2}\frac{{d{{\left| {{V_{ij}}} \right|}^2}}}{{dt}} &=& S(t)\Big\{ 3{\left| {{V_{ij}}} \right|^2}({f_i}^2 + {h_j}^2 - \sum\limits_k {{f_k}^2} {\left| {{V_{kj}}} \right|^2} - \sum\limits_k {{h_k}^2} {\left| {{V_{ik}}} \right|^2}) \nonumber\\
&& - 3{f_i}^2\sum\limits_{k \ne i} {\frac{1}{{{f_i}^2 - {f_k}^2}}} (2{h_j}^2{\left| {{V_{kj}}} \right|^2}{\left| {{V_{ij}}} \right|^2} + \sum\limits_{l \ne j} {h_l}^2{V_{iklj}})\nonumber\\
&& - 3{h_j}^2\sum\limits_{l \ne j} {\frac{1}{{{h_j}^2 - {h_l}^2}}} (2{f_i}^2{\left| {{V_{il}}} \right|^2}{\left| {{V_{ij}}} \right|^2} + \sum\limits_{k \ne i} {f_l}^2{V_{iklj}}) \Big\},  \label{eqn:9}
\eea
where ${V_{iklj}} = 1 - {\left| {{V_{il}}} \right|^2} - {\left| {{V_{kl}}} \right|^2} - {\left| {{V_{kj}}} \right|^2} - {\left| {{V_{ij}}} \right|^2} + {\left| {{V_{il}}} \right|^2}{\left| {{V_{kj}}} \right|^2} + {\left| {{V_{kl}}} \right|^2}{\left| {{V_{ij}}} \right|^2}$, and the structure of the one-loop evolution equation for the gauge couplings are given by
\beq
16{\pi ^2}\frac{{d{g_i}}}{{dt}} = [{b_i}^{SM} + (S(t) - 1){{b}_i^{UED}}]{g_i}^3 \; . \label{eqn:10}
\eeq
The ${b_i}^{SM} = (\frac{{41}}{10}, - \frac{{19}}{6}, - 7)$ and ${{b}_i^{UED}} = (\frac{{81}}{{10}},\frac{7}{6}, - \frac{5}{2})$. Thus, Eqs.(\ref{eqn:5},\ref{eqn:8},\ref{eqn:9},\ref{eqn:10}) form a complete set of coupled differential equations for the three families.


\section{Precise Solutions of the Higgs Self-Coupling Equation}\label{sec:3}

\par The evolution equations relate various observables at different energy scales, and which also allow one to study the asymptotic or perturbative behaviors of these equations at a higher energy scale. The one-loop evolution equation, Eq.(\ref{eqn:5}), for the Higgs quartic coupling $\lambda$ is nonlinear, which can be rewritten as:
\bea
\frac{{d\lambda }}{{dt}} &=& \frac{1}{{16{\pi ^2}}}\{ 12 \cdot S(t){\lambda ^2} - S(t) \cdot \left( {\frac{9}{5}g_1^2 + 9g_2^2} \right)\lambda + (2S(t) - 1) \cdot 4Tr[3Y_U^\dag {Y_U} + 3Y_D^\dag {Y_D} + Y_E^\dag {Y_E}]\lambda \nonumber \\
&& + \frac{9}{4}\left( {\frac{3}{{25}}g_1^4 + \frac{2}{5}g_1^2g_2^2 + g_2^4} \right) + (S(t) - 1) \cdot \left( {\frac{9}{{25}}g_1^4 + \frac{6}{5}g_1^2g_2^2 + 3g_2^4} \right) \nonumber \\
&& - (2S(t) - 1) \cdot 4Tr[3{(Y_U^\dag {Y_U})^2} + 3{(Y_D^\dag {Y_D})^2} + {(Y_E^\dag {Y_E})^2}]\} \nonumber \\
& =& {f_0}(t) + {f_1}(t)\lambda  + {f_2}(t){\lambda ^2} \; , \label{eqn:11}
\eea
where
\bea
{f_0}(t) &=& \frac{1}{{16{\pi ^2}}}\{ \frac{9}{4}\left( {\frac{3}{{25}}g_1^4 + \frac{2}{5}g_1^2g_2^2 + g_2^4} \right) + (S(t) - 1) \cdot \left( {\frac{9}{{25}}g_1^4 + \frac{6}{5}g_1^2g_2^2 + 3g_2^4} \right) \nonumber \\
&& - (2S(t) - 1) \cdot 4Tr[3{(Y_U^\dag {Y_U})^2} + 3{(Y_D^\dag {Y_D})^2} + {(Y_E^\dag {Y_E})^2}]\}, \nonumber \\
{f_1}(t) &=& \frac{1}{{16{\pi ^2}}}\{  - S(t) \cdot \left( {\frac{9}{5}g_1^2 + 9g_2^2} \right) + (2S(t) - 1) \cdot 4Tr[3Y_U^\dag {Y_U} + 3Y_D^\dag {Y_D} + Y_E^\dag {Y_E}]\}, \nonumber \\
{f_2}(t) &=& \frac{1}{{16{\pi ^2}}}12 \cdot S(t) \; . \label{eqn:12}
\eea
Explicitly Eq.(\ref{eqn:11}) has the form of the Riccati differential equation\cite{Kielanowski:2003jg}, which enables us to solve the equation explicitly. In fact, the solutions of Riccati's equation can become singular even if the coefficients ${f_0}(t)$, ${f_1}(t)$, and ${f_2}(t)$ of the equation are smooth and regular functions of energy. The position of the singularities and zeros of $\lambda (t)$ can be determined precisely, and their dependence on the initial value of the Higgs quartic coupling $\lambda (M_Z)$ can also be derived. Consider two independent solutions ${W_1}(t)$ and ${W_2}(t)$ which satisfy the following differential equation:
\beq
W'' - \left(\frac{{{f_2}'(t)}}{{{f_2}(t)}} + {f_1}(t)\right)W' + {f_0}(t){f_2}(t)W = 0 \; , \label{eqn:13}
\eeq
along with the initial conditions ${W_1}({t_0}) = 1$, ${W_1}'({t_0}) = 0$, ${W_2}({t_0}) = 0$, and ${W_2}'({t_0}) = 1$.  In terms of the functions ${W_1}(t)$ and ${W_2}(t)$ the solution for $\displaystyle \lambda (t)= -\frac{{1}}{f_2(t)} \frac{{{W}'(t)}}{{ W(t)}}$ thus gives us
\beq
\lambda (t) =  - \frac{{16{\pi ^2}}}{{12S(t)}}\frac{{{W_1}'(t) - \displaystyle \frac{{12}}{{16{\pi ^2}}}S(t = {t_0})\lambda (t = {t_0}){W_2}'(t)}}{{{W_1}(t) - \displaystyle \frac{{12}}{{16{\pi ^2}}}S(t = {t_0})\lambda (t = {t_0}){W_2}(t)}}\; , \label{eqn:14}
\eeq
where ${t_0} = \ln (\frac{1}{{{M_Z}R}})$ is the place at which the first KK level is excited. Note that $S(t = {t_0}) = 1$, and the initial value $\lambda (t = {t_0})$ is to be determined. Obviously, the singularity and the zero of the solution $\lambda (t)$ depend on the compactification radius $R$ and can be read off directly from the zeros of the denominator and numerator respectively. In which case the singularity condition leads to
\beq
{\lambda _S}(t = {t_0}) = \frac{{{W_1}(t)}}{{ \displaystyle \frac{{12}}{{16{\pi ^2}}}{W_2}(t)}}\; . \label{eqn:15}
\eeq
That is, for a given initial value ${\lambda _S}(t = {t_0})$ the evolution of $\frac{{{ \displaystyle W_1}(t)}}{{ \displaystyle \frac{{12}}{{16{\pi ^2}}}{W_2}(t)}}$ gives us a Landau pole of $\lambda (t)$ when it equals to ${\lambda _S}(t = {t_0})$ at an energy cutoff $t = \Lambda$. Similarly, as observed from the numerator, if we start off from a different value ${\lambda _{Z}}(t = {t_0})$, the evolution of $\frac{{{ \displaystyle W_1}'(t)}}{{ \displaystyle \frac{{12}}{{16{\pi ^2}}}{W_2}'(t)}}$ would give us a zero value of $\lambda (t)$ at a certain energy scale. In which case we have the following equality
\beq
{\lambda _Z}(t = {t_0}) = \frac{{{W_1}'(t)}}{{ \displaystyle \frac{{12}}{{16{\pi ^2}}}{W_2}'(t)}}\; . \label{eqn:16}
\eeq

\par
\begin{figure}[tb]
\begin{center}
\epsfig{file=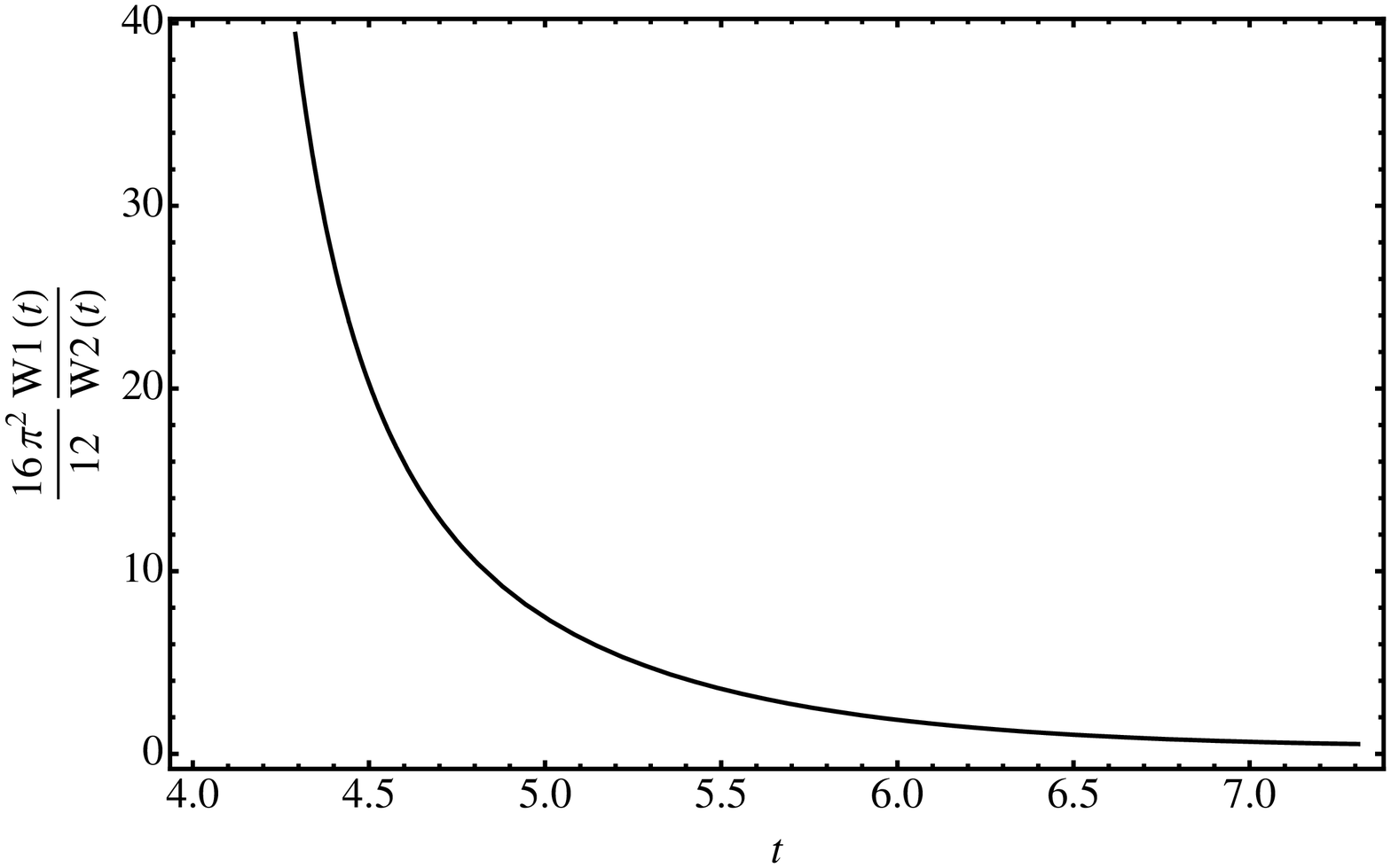,width=.4\textwidth}
\caption{\sl The evolution of $\frac{{{W_1}(t)}}{{\frac{{12}}{{16{\pi ^2}}}{W_2}(t)}}$ from $t_0$ to the unification scale for ${R^{ - 1}} = 5TeV$}
\label{fig:2}
\epsfig{file=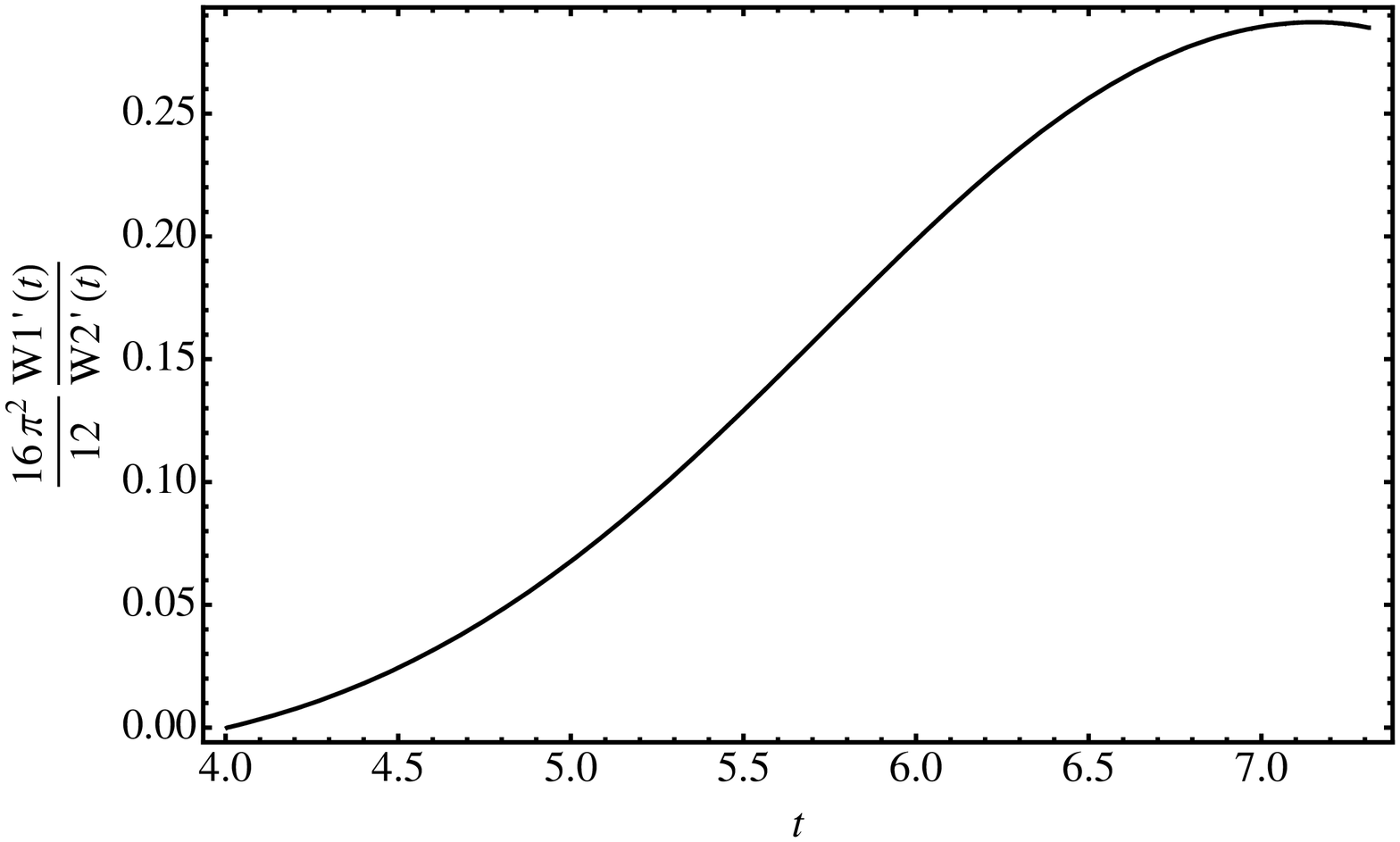,width=.4\textwidth}
\caption{\sl The evolution of $\frac{{{W_1}'(t)}}{{\frac{{12}}{{16{\pi ^2}}}{W_2}'(t)}}$ from $t_0$ to the unification scale for ${R^{ - 1}} = 5TeV$}
\label{fig:3}
\end{center}
\end{figure}

 In Figs.\ref{fig:2} and \ref{fig:3} we plot $\frac{{{ \displaystyle W_1}(t)}}{{ \displaystyle \frac{{12}}{{16{\pi ^2}}}{W_2}(t)}}$ and $\frac{{{ \displaystyle W_1}'(t)}}{{ \displaystyle \frac{{12}}{{16{\pi ^2}}}{W_2}'(t)}}$ as functions of the scale parameter $t$. In which, for definiteness, we choose ${R^{ - 1}} = 5TeV$ as an illustrative example, and which is within the reach of the LHC. Note that when the energy is greater than $5 TeV$ we can see from Eq.(\ref{eqn:11}) that the beta function is governed by the whole SM sector, as well as its KK counterpart. As illustrated, the function $\frac{{{ \displaystyle W_1}(t)}}{{ \displaystyle \frac{{12}}{{16{\pi ^2}}}{W_2}(t)}}$ decreases monotonically with energy, and it approaches the value 0.537 at the unification scale, where the gauge couplings tend to converge. Therefore, for the initial value $\lambda_S (t = {t_0})  =0.537$, the $\lambda(t)$ eventually develops a Landau pole and ``blows-up''. If we require the ``blow-up'' should not happen at an energy scale smaller than the unification scale, $\lambda_S (t = {t_0})$ must be no more than 0.537. Furthermore, for an energy below the threshold of the first KK level, we can track back to the initial value of $\lambda(t)$ at the electroweak scale by using the SM beta function in Eq.(\ref{eqn:5}), thus we can determine $\lambda (M_Z)$. As a result, in terms of the Higgs mass, for ${R^{ - 1}} = 5TeV$, it constraints ${m_H}({M_Z}) <184.1GeV$ if the theory is valid in the range from the electroweak scale up to the unification scale.

\par On the other hand, the self-interactions of the scalar field should remain in the perturbative domain, and no instabilities should develop in the whole energy range between the electroweak scale and the unification scale. In another words, the running of the self-coupling $\lambda$ is required to remain positive between the electroweak scale and the ultraviolet cutoff. In Fig.\ref{fig:3} we find the function $\frac{{{ \displaystyle W_1}'(t)}}{{ \displaystyle \frac{{12}}{{16{\pi ^2}}}{W_2}'(t)}}$ keeps increasing from the scale where the first KK mode is excited up to the unification scale. However, as observed from the figure, the increase is not monotonical. Before the function gets to the final value, it reaches a maximum value of 0.287, just a marginally lower than the unification scale. Like any higher dimensional theory, the UED model should be treated only as an effective theory which is valid only up to some scale $\Lambda$, at which a new physics theory emerges. Therefore, presumably the new physics will be associated with the scale where the function $\frac{{{ \displaystyle W_1}'(t)}}{{ \displaystyle \frac{{12}}{{16{\pi ^2}}}{W_2}'(t)}}$ becomes maximum.  For if the initial value of $\lambda ({t_0})$ were less than 0.287, the evolution of $\lambda (t)$ would pass through zero and become negative on its way to the unification scale, in which case the Higgs sector will break down and the theory would become invalid. Thus the energy scale where $\lambda(t)$ becomes negative defines a new scale, and in the current context it is found to be different from the unification scale, which is in contrast with the results of the pure SM \cite{Kielanowski:2003jg}.  After we trace back to the $\lambda$ value at the electroweak scale, by using the beta function of the SM for energies below ${R^{ - 1}}$, we obtain a lower bound on the Higgs mass, i.e. ${m_H}({M_Z})>154.4GeV$, if one requires that the vacuum is not destabilized up to the unification scale.


\section{Numerical Analysis}\label{sec:4}

\par
\begin{figure}[tb]
\begin{center}
\epsfig{file=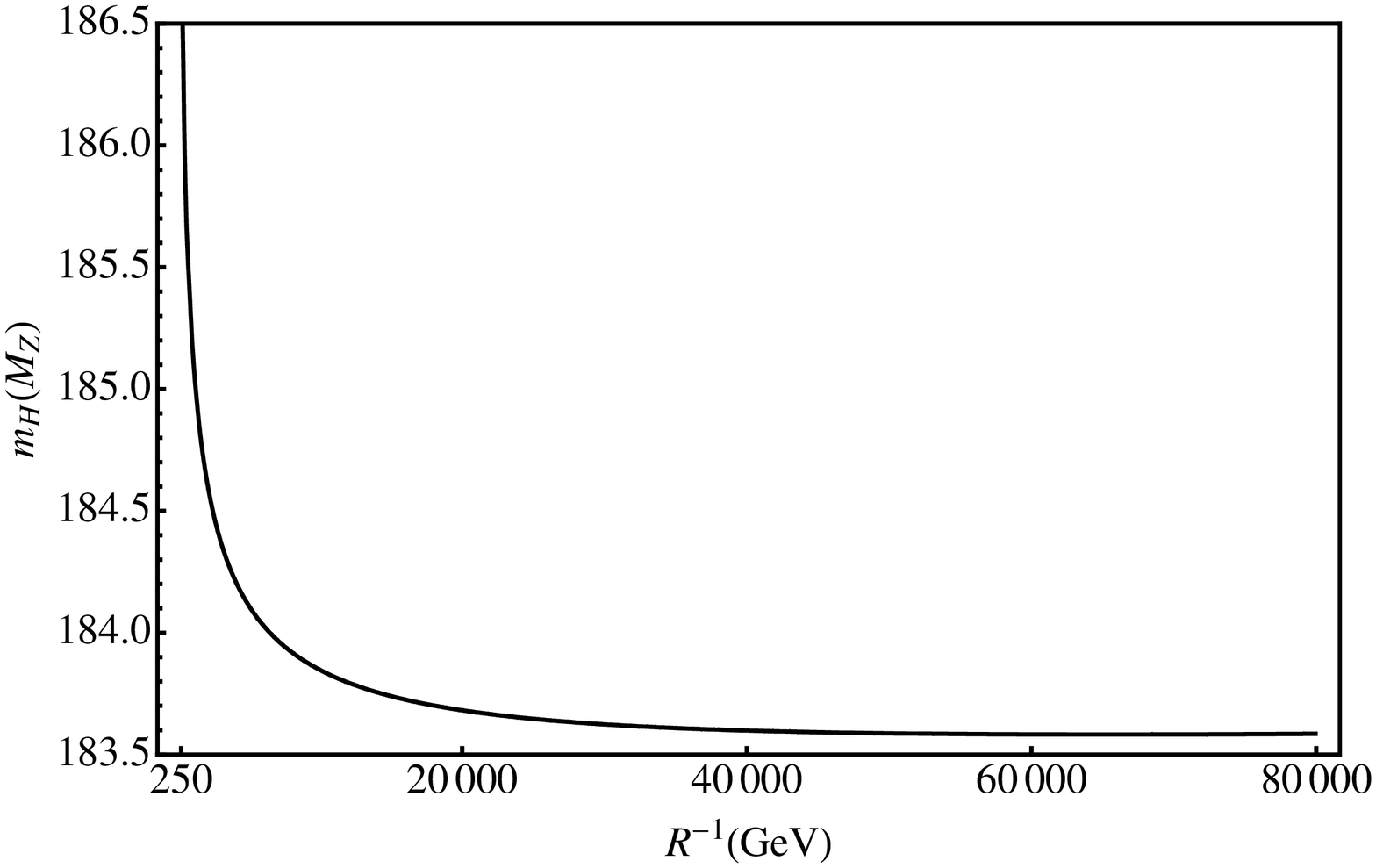,width=.4\textwidth}
\caption{\sl The one-to-one correspondence between the Higgs mass and the compactification scale $R^{-1}$, derived from the singularity condition of the scalar coupling $\lambda (t)$.}
\label{fig:4}
\epsfig{file=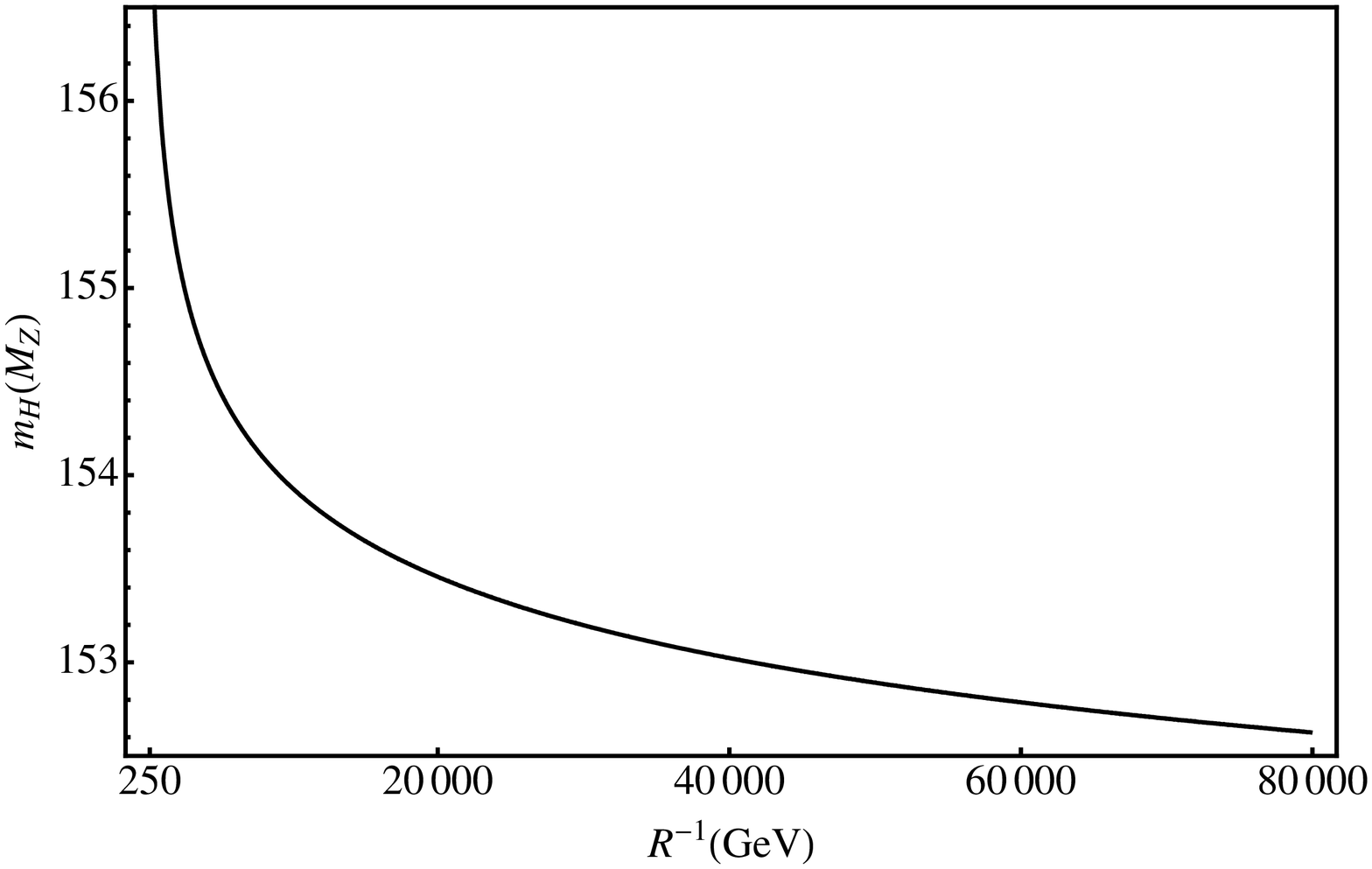,width=.4\textwidth}
\caption{\sl The one-to-one correspondence between the Higgs mass and the compactification scale $R^{-1}$, derived from the vacuum stability condition of the scalar coupling $\lambda (t)$.}
\label{fig:5}
\end{center}
\end{figure}

 We shall now investigate how the evolution of $\lambda$ depends on the initial values of $\lambda(M_Z)$ and thus the constraints imposed on the compactification radii for which the validity of the theory is satisfied. Also, as is well known, the self-coupling of the Higgs fields $\lambda(M_Z)$ is proportional to the Higgs mass squared, i.e., $\lambda  = m_H^2/{v^2}$; the admissible values of $\lambda(M_Z)$ can be transformed into the allowed values of the Higgs boson mass at $M_Z$. For different compactification radii $R$ we can apply Eqs.(\ref{eqn:14},\ref{eqn:15},\ref{eqn:16}) to fix the value of $\lambda(t)$ at ${t_0} = \ln (\frac{1}{{{M_Z}R}})$ using the singularity and vacuum stability conditions of the scalar coupling. The initial value of $\lambda$ at $t=0$ is then determined by the beta function of the SM in Eq.(\ref{eqn:5}). In the range of compactification radius $250GeV \sim {R^{ - 1}} \sim 80TeV$ we have plotted the one-to-one correspondence of the compactification radius and the initial Higgs mass $m_H(M_Z)$ for both the singularity and zero conditions of $\lambda(t)$. In Fig.\ref{fig:4} a given point on the curve corresponds to the compactification radius $R^{-1}$ and the associated maximum initial value of the Higgs mass(where we have reformulated the initial values of $\lambda ({M_Z})$ to the Higgs mass $m_H$), and any Higgs mass larger than that will lead to the divergence of $\lambda(t)$ before the unification scale. Similarly, for Fig.\ref{fig:5} a point on the curve represents the lower limit of the Higgs mass for the specific $R^{-1}$, and any Higgs mass lower than that will lead to the evolution of $\lambda(t)$ becoming negative before reaching the unification scale.

\par
\begin{figure}[tb]
\begin{center}
\epsfig{file=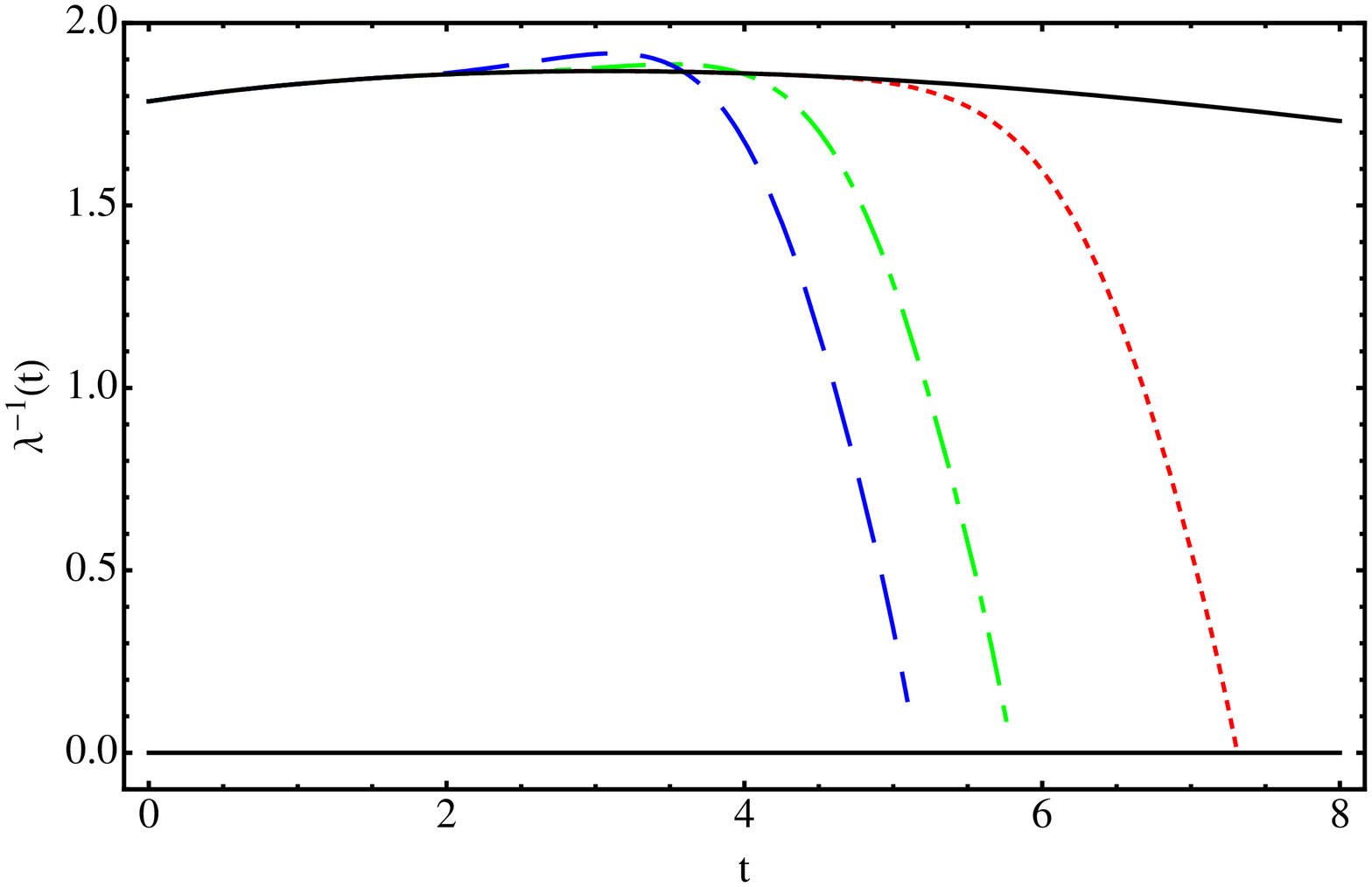,width=.4\textwidth}
\caption{\sl Graph for the evolution of $\lambda^{-1}(t)$ related to the singularity condition. Here, $\lambda(M_Z)=0.560$, and dotted line is the $R^{-1}=5TeV$ UED case, which reaches zero at the unification scale. The dotted-dashed line is the $R^{-1}=1TeV $ UED case, the dashed line is the $R^{-1}=500GeV$ UED case, and the solid line is the SM.}
\label{fig:6}
\epsfig{file=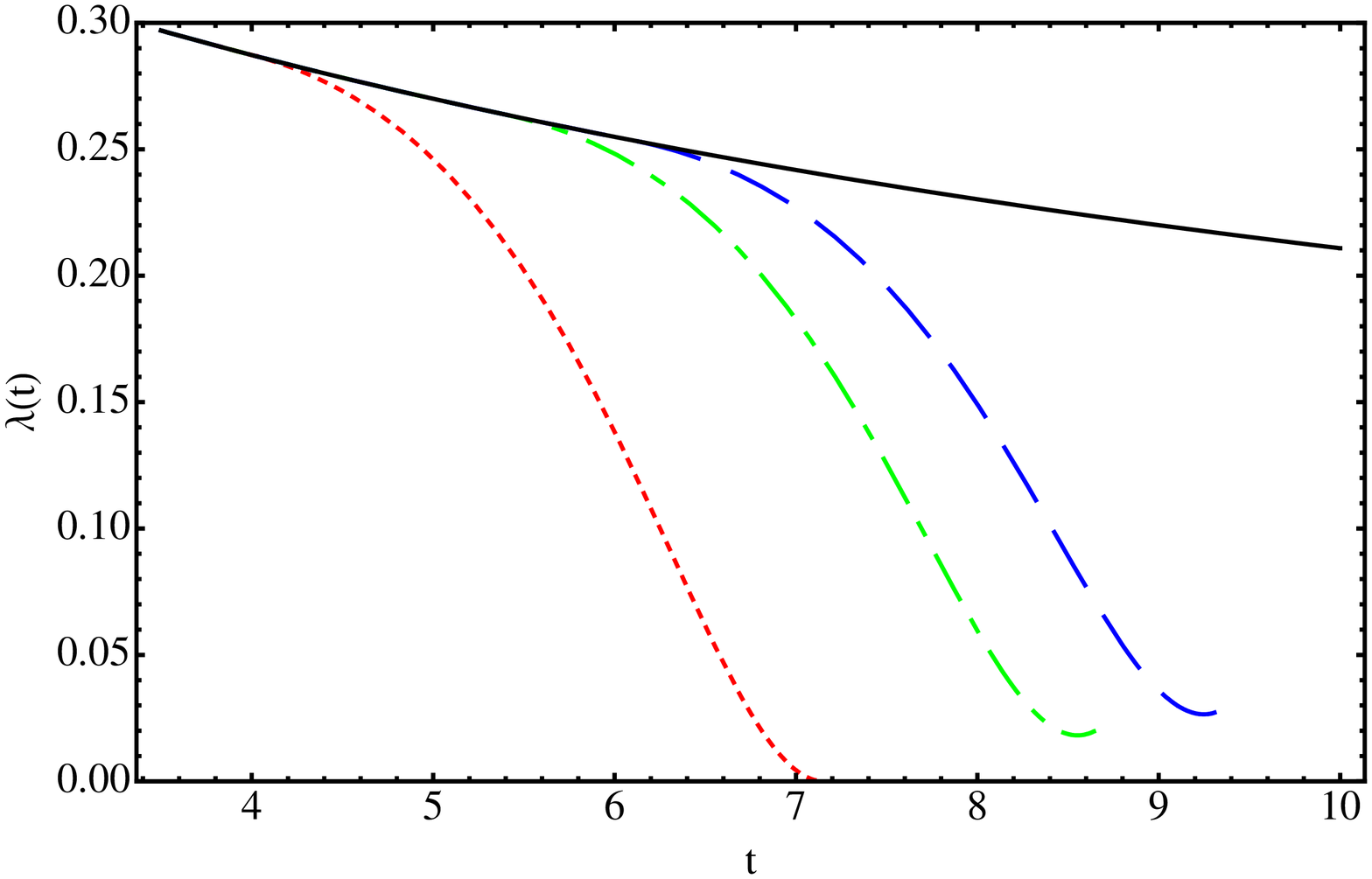,width=.4\textwidth}
\caption{\sl Graph for the evolution of $\lambda(t)$ related to the vacuum stability condition. Here, $\lambda(M_Z)=0.394$, where the dotted line is the $R^{-1}=5TeV$ UED case, the dotted-dashed line is the $R^{-1}=20TeV$ UED case, the dashed line is the $R^{-1}=40TeV$ UED case, and the solid line is the SM.}
\label{fig:7}
\epsfig{file=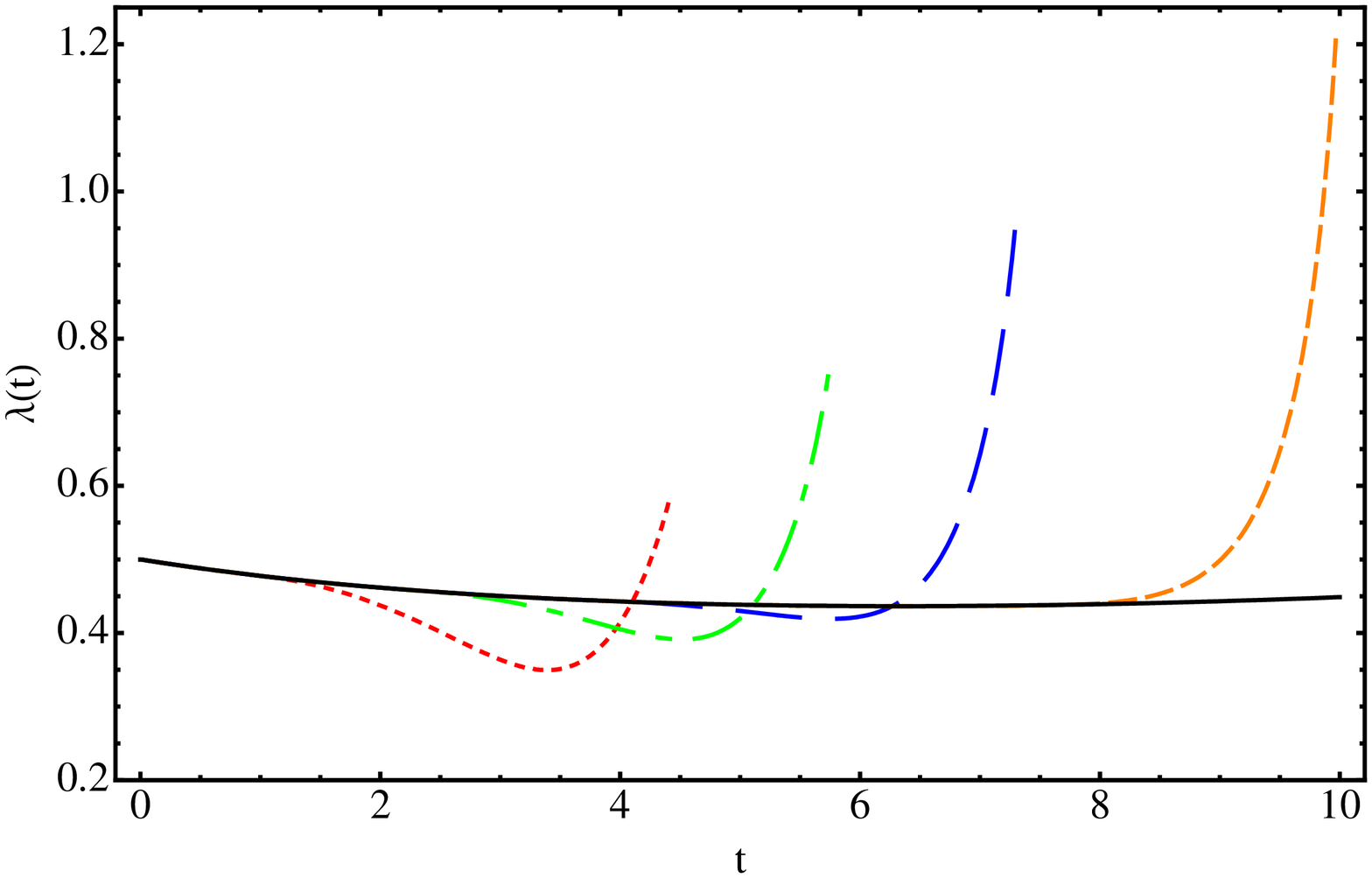,width=.4\textwidth}
\caption{\sl Graph for the evolution of $\lambda(t)$ related to the intermediate value of $\lambda(M_Z)$. Here, $\lambda(M_Z)=0.50$, where the dotted line is the $R^{-1}=250GeV$ UED case, the dotted-dashed line is the $R^{-1}=1TeV$ UED case, the long dashed line is the $R^{-1}=5TeV$ UED case, the short dashed line is the $R^{-1}=80TeV$ UED case, and the solid line is the SM.}
\label{fig:8}
\end{center}
\end{figure}

 Furthermore, for a given initial value of $\lambda(t)$ we can follow the differential equation Eq.(\ref{eqn:5}) to pursue its evolution for different compactification radii. For definiteness, in Figs.\ref{fig:6} and \ref{fig:7} we plot the energy dependence of $\lambda (t)$ for different compactification radii, and thus illustrate its bounds on ${R^{ - 1}}$.\footnote{Here we plotted the $\lambda(t)$ evolution for the UED model using Eq.(\ref{eqn:5}) and checked it with Eq.(\ref{eqn:14}) and found they agreed extremely well.} As depicted in Fig.\ref{fig:6}, when the energy of the system is less than the excitations of the first KK modes the theory follows the evolution of the usual 4-dimensional SM, and the existence of the KK modes is thus ignored. However, once the first KK threshold is reached when $\mu  > {R^{ - 1}}$, the contributions from the KK states become more and more significant, and the running deviates from its normal SM orbits and begins to evolve with a faster rate. For the initial value of $\lambda ({M_Z}) = 0.560$, we find that the evolution of the coupling $\lambda (t)$ develops a Landau pole at the unification scale for compactification radius ${R^{ - 1}} = 5TeV$ (i.e., the inverse of $\lambda (t)$ becomes zero in Fig.\ref{fig:6}, and this result can also be concluded from Sec.\ref{sec:3}. For other radii ${R^{ - 1}}$ that are less than $5TeV$, for example, ${R^{ - 1}} = 500GeV$, and $1TeV$ as shown in Fig.\ref{fig:6}, there is no Landau pole up to the unification scale and the evolution of the coupling $\lambda (t)$ is finite (the unification scale is chosen to be the median point of the area where the gauge couplings converge); however, any theories whose radius ${R^{ - 1}} > 5TeV$ are ruled out in this context as the evolution of $\lambda (t)$ becomes divergent and singular before reaching the unification scale.

\par In Fig.\ref{fig:7} we show an alternative evolution scenario for $\lambda (t)$ when we start with a different initial value of $\lambda ({M_Z})$ in the range of a Higgs mass of $152GeV \sim {m_H}({M_Z}) \sim 157GeV$.  In fact, for a small ${m_H}({M_Z})$ and a large ${({Y^\dag }Y)^2}$ term, the last term in Eq.(\ref{eqn:11}) dominates and has a negative contribution to the beta function. This causes the scalar self coupling to decrease with scale. Since the Yukawa coupling itself falls with scale, especially in the UED model, the Yukawa couplings are driven dramatically towards extremely weak values at a much a faster rate\cite{Cornell:2010sz}. Thus, eventually, the ${\lambda ^2}$ term in Eq.(\ref{eqn:11}) will overwhelm and overcome the negative contribution from the Yukawa couplings and bring the beta function back to positive values. Qualitatively, the function $\lambda (t)$ will initially fall with scale until a minimum is reached, and then rise. If this minimum is above zero, the vacuum of the theory is stable. However, if the minimum becomes negative, this will destabilize the vacuum. In Fig.\ref{fig:7} we start from $\lambda ({M_Z}) =0.394$ and follow Eq.(\ref{eqn:5}) to track its evolution for different compactification radii. As concluded from Sec.\ref{sec:3}, for ${R^{ - 1}} = 5TeV$, the scalar coupling reaches zero at a certain energy scale. In order to rescue the stability of the vacuum, it suggests it is necessary to introduce new physics at such a scale which would have a non-negligible impact on the radiative corrections to the scalar potential and raise it. Therefore, at or below this scale a consistent description of nature requires the introduction of new physics. For other compactification radii ${R^{ - 1}} = 10TeV$, and $20TeV$, for example, as observed in Fig.\ref{fig:7}, the evolution of $\lambda (t)$ reaches its minimum but rises and remains positive in the whole range from the electroweak scale to the unification scale. As a result, for initial value $\lambda ({M_Z}) =0.394$, it rules out any compactification radii ${R^{ - 1}}$ that are less than $5TeV$ for theories that are valid and whose vacuum is stable against radiative corrections up to the unification scale. In Fig.\ref{fig:8}, we plot the evolution of the $\lambda (t)$ for the initial value $\lambda ({M_Z}) =0.50$ located in the range of the Higgs mass $157GeV < {m_H}({M_Z}) < 183GeV$. It is shown that, for compactification radii $250GeV \sim {R^{ - 1}} \sim 80TeV$, the scalar coupling is positive and non-singular from the electroweak scale up to the unification scale. This can also be expected from the results of Figs.\ref{fig:4} and \ref{fig:5}, since this initial value of the scalar coupling is outside the singularity and vacuum stability constraints for  compactification radii $250GeV \sim {R^{ - 1}} \sim 80TeV$.

\section{Summary}\label{sec:5}

\par The Higgs sector provides us with a number of interesting problems in particle physics, where in this paper we have investigated the evolution of the scalar coupling for the UED model, and analyzed its effects and patterns for different initial values of $\lambda ({M_Z})$ and different compactification radii $R$. Moreover, an analytical and numerical solution of the one-loop evolution of the Higgs quartic coupling $\lambda$ is obtained, where the analysis of the one-loop equation gives explicit formulae for the singularity and zero positions of the scalar coupling. For compactification radius between $250GeV \sim {R^{ - 1}} \sim 80TeV$, by means of singularity and vacuum stability conditions, an intimate correspondence and connection between the Higgs mass and the compactification radii ${R^{ - 1}}$ is explicitly plotted.

\par If we consider a consistent theory, one where the running of the scalar coupling remains positive and non-singular in the whole range of energies between the electroweak scale and the unification energy, we have constrained the range for the Higgs mass in the UED model and are able to limit the range of compactification radius for different initial values of $\lambda (M_Z)$. For a compactification radius in the range between $250GeV \sim {R^{ - 1}} \sim 80TeV$, and for a large Higgs mass, it gives us an upper bound on the compactification scale $R^{-1}$, where any other compactification scales beyond that will develop a Landau pole before the unification scale is reached. Therefore, the theory will have a strong coupling at some high scale and will no longer be a complete or consistent theory to describe the model. If we start with a light Higgs mass instead, requiring $\lambda$ remain positive up to the unification scale, the evolution of the scalar coupling will lead to a lower bound on the compactification scale $R^{-1}$. Below this compactification scale the theory will not be valid in the whole range from the electroweak scale to the unification scale, since the evolution of $\lambda$ becomes negative and destabilizes the vacuum. It is thus expected that new physics should come to the fore in order to raise the potential. If the Higgs mass is between $152GeV \sim {m_H}({M_Z}) \sim 157GeV$, then there is no vacuum stability and singularity concerns all the way up to the unification scale, yielding no bounds on the compactification radius $R$.  Any other Higgs mass that is outside the range illustrated here is usually described by a finite cut-off scale, where the model breaks down and new physics appears. If the compactification radius $R$ is sufficiently large, due to the power law running of the gauge couplings, this enables us to bring the unification scale down to an exportable range at the LHC scale. Therefore, these bounds are very relevant for us, especially for a compactification scale that lies in a region which could be accessible for the LHC or future accelerators. Hence it is expected that our conclusions shall set a limit on the UED model to satisfy both theoretical and experimental consistency.

\end{document}